\documentclass[sigconf]{acmart}

\usepackage{multirow}
\usepackage{subfigure}
\usepackage{bm}
\usepackage{balance}
\setcopyright{none}

\acmConference[Accepted by SIGIR]{}{2020}{10 pages}

\acmBooktitle{Proceedings of the 43rd International ACM SIGIR Conference on Research and Development in Information Retrieval (SIGIR '20), China}

\author{Kelong Mao}
\affiliation{Tsinghua University}
\affiliation{Huawei Noah's Ark Lab}
\email{mkl18@mails.tsinghua.edu.cn}

\author{Xi Xiao}
\affiliation{Tsinghua University}
\affiliation{Pengcheng Lab}
\email{xiaox@sz.tsinghua.edu.cn}

\author{Jieming Zhu}
\authornote{Corresponding author}
\affiliation{Huawei Noah's Ark Lab}
\email{jamie.zhu@huawei.com}

\author{Biao Lu}
\affiliation{Huawei Noah's Ark Lab}
\email{lubiao4@huawei.com}

\author{Ruiming Tang}
\affiliation{Huawei Noah's Ark Lab}
\email{tangruiming@huawei.com}

\author{Xiuqiang He}
\affiliation{Huawei Noah's Ark Lab}
\email{heixiuqiang1@huawei.com}

\begin{document}
\fancyhead{}

\title{Item Tagging for Information Retrieval: A Tripartite Graph Neural Network based Approach}


\begin{abstract}
Tagging has been recognized as a successful practice to boost relevance matching for information retrieval (IR), especially when items lack rich textual descriptions. A lot of research has been done for either multi-label text categorization or image annotation. However, there is a lack of published work that targets at item tagging specifically for IR. Directly applying a traditional multi-label classification model for item tagging is sub-optimal, due to the ignorance of unique characteristics in IR. In this work, we propose to formulate item tagging as a link prediction problem between item nodes and tag nodes. To enrich the representation of items, we leverage the query logs available in IR tasks, and construct a query-item-tag tripartite graph. This formulation results in a TagGNN model that utilizes heterogeneous graph neural networks with multiple types of nodes and edges. Different from previous research, we also optimize both full tag prediction and partial tag completion cases in a unified framework via a primary-dual loss mechanism. Experimental results on both open and industrial datasets show that our TagGNN approach outperforms the state-of-the-art multi-label classification approaches. 

\end{abstract}

%



\maketitle

\section{Introduction}
\begin{figure}[!t]
  \centering
  \includegraphics[width=0.35\textwidth]{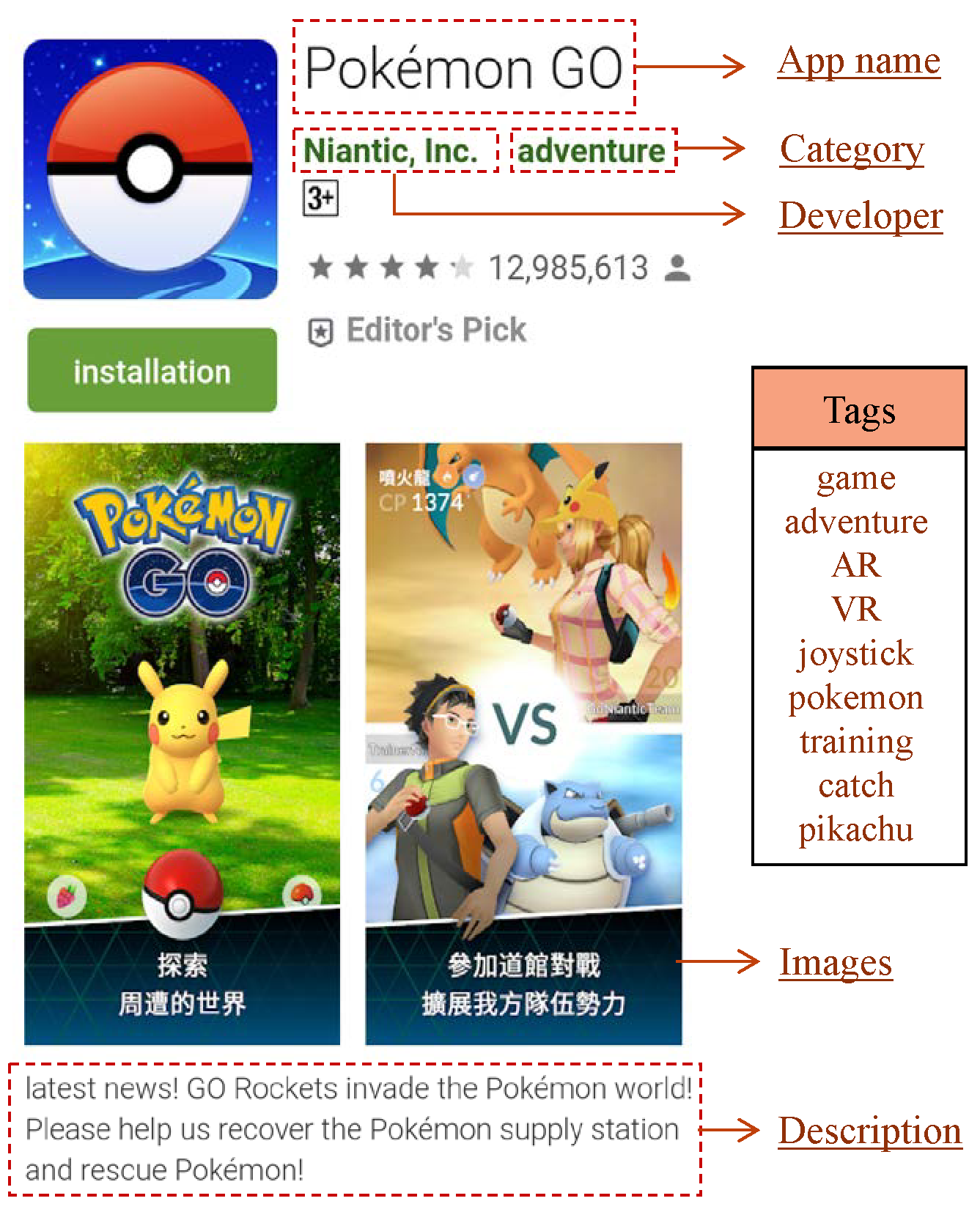}
  \caption{An Illustrative Item Example
}\label{fig:app_demo}
\end{figure}
Information retrieval (IR) is a well-established research area that deals with our daily information needs, such as Web search, App search, e-commence product search, image retrieval, music finding, and so on. Although text-based Web search has been widely studied in the literature, IR in vertical domains faces some unique challenges. Different from Web search that mostly deals with full-text documents, textual descriptions of items in some other domains are not sufficiently rich or concise to convey their semantic information. For illustration, we take app search as an example. Figure~\ref{fig:app_demo} presents an example app (i.e., Pokemon Go) from Google Play. It consists of multiple types of information including app name, category, developer, screenshot images, and a short description. The description, however, comprises a promotion news only. Such short and noisy item descriptions increases the difficulty for retrieving relevant items.

In such scenarios, tagging plays a critical role in helping describe and enrich the semantics of items. Tags are often characterized as keywords to describe the key information of items such as category, functionality, style, related entities, target audience, etc. Tagging has been recognized as a successful practice to boost the retrieval performance, especially for those items that lack concise textual descriptions~\cite{Tag-IR-survey}. For instance, the app item in Figure~\ref{fig:app_demo} has a set of tags including "game", "AR" (Augmented Reality), "pikachu", etc. These tags make it easier to retrieval the app when a user searches the query "pikachu game" or "AR game", but this cannot be done from the textual description only. The collection of tags can not only boost relevance matching, but also be used for query reformulation and item recommendation~\cite{Tag_rec}. In addition, displaying tags and clickable hyperlinks along with their associated items can help users navigate and explore item collections of interest.

For many industrial IR applications, item tagging serves as a key building block for better item organization and retrieval. For user generated content, tags are provided by users themselves for their posts (e.g., tweets hashtags in Twitter, question tags in StackOverflow). In contrast, for platform generated content (i.e., items), such as apps, ads and news, tags and their integration to search may be not visible to users. Item tagging becomes a regular task of operation teams~\cite{manual_tagging}. However, manual tagging is a time-consuming process and might result in unmanageable efforts when the item corpus is too large. To replace or supplement the manual tagging process, a large body of research has been done toward automatic item tagging. Typical examples include app tagging~\cite{app_tagging}, news tagging~\cite{news_tagging_1,news_tagging_2}, blog posts tagging~\cite{post_tagging_1, post_tagging_2}, questions tagging~\cite{question_tag_1,question_tag_3}, image annotation~\cite{image_tagging_1,image_tagging_3}. 

Potential methods for item tagging can be broadly categorized into two types: \textit{keyphrase extraction}~\cite{keyprhase_survey} and \textit{multi-label classification}~\cite{MTL_survey}. Keyphrase extraction methods (e.g., TF-IDF~\cite{keyphrase_review}, TextRank~\cite{TextRank}, PositionRank~\cite{PositionRank}) have been widely used for textual documents or websites to identify keywords from original content that best describe the subject of a document. These methods mostly follow a two phase procedure (i.e., candidate extraction $\rightarrow$ ranking). They work well for long documents but are inappropriate for items without detailed textual descriptions, because tags might not appear in the item description. As such, item tagging is often formulated as a multi-label classification problem~\cite{MTL_survey}, that is, assigning relevant tags to items from a collection of predefined ones. Multi-label classification models have been widely studied in the literature, and many of them are successfully applied to text categorization~\cite{MTL_text1,MTL_text2,MTL_text3}. 
However, directly applying a traditional multi-label classification model for item tagging is sub-optimal, especially in information retrieval tasks.

In this work, inspired by the recent success of graph neural networks (GNN)~\cite{GNN_survey}, we propose to cast item tagging as a link prediction problem between item nodes and tag nodes, and present a GNN-based model for item tagging (namely TagGNN). In contrast to previous research, our work aims to address the following limitations.
\begin{itemize}
\item Most traditional multi-label classification models cannot fully exploit the correlations among tags (i.e., labels). Instead, our formulation enables tag embedding via node representation, which better captures the correlations among similar tags. It also enriches the representation of item and tag nodes, since semantically similar information can be aggregated from neighbour nodes via message passing. Intuitively, items and tags are matched not only by themselves, but also by neighbour items and neighbour tags. 

\item Item descriptions are usually short and noisy, making it difficult to extract semantic information from textual descriptions for classification. To alleviate this issue, we propose to not only utilize the textual descriptions, but also leverage the query logs available to enrich the representation of items. We construct a query-item-tag tripartite graph, where query-item edges indicate the interactions (e.g., clicks or downloads) in the query log and item-tag edges represent the annotation relationships. This tripartite graph is unique for IR and leads to heterogeneous GNN modeling with multiple types of nodes and edges. Our TagGNN model naturally fuses item-tag (w.r.t. TagGNN-IT) and query-item (w.r.t. TagGNN-QI) graphs.

\item In practice, some new items have no existing tags and need to make full tag prediction. Some old items have partial incomplete tags (e.g., manually labelled), which only need tag completion and refinement. Both cases are desired in IR tasks. While existing work focuses on either one~\cite{MTL_text3} or the other~\cite{POI_tagging}, we optimize both cases in a unified framework. To achieve this, we join a primary loss and a dual loss during training to avoid training-testing exposure bias.

\end{itemize}

We also emphasize that, while some work that leverages GNNs for text categorization exists~\cite{GNN_text1,GNN_text2,GNN_text3}, we are not aware of any published work about GNN-based item tagging that is formulated as a link prediction problem. To evaluate the effectiveness of our TagGNN approach, we conduct comprehensive experiments on two large datasets, including an open dataset of ad tagging for sponsored product search ({KDDCup-2012}) and a private industrial app tagging dataset for app search ({Huawei-Dataset}). The experimental results show that our TagGNN approach achieves consistent improvements in precision over 9 baseline models in both "without tags" and "partial tags" settings. Ablation studies and parameter analyses have also been conducted to validate our model design choices. 

In summary, our work makes the following main contributions:
\begin{itemize}
    \item Our work formulates item tagging as a link prediction problem over the query-item-tag graph and present a unique tripartite-graph neural network based approach.
    \item We target at both full tag prediction and partial tag completion, and present a primary-dual losses to optimize both cases in a unified learning framework.
    \item Our experimental results show significant improvements over both text-based and graph-based competing methods. 
\end{itemize}

The remainder of this paper is organized as follows. Section~\ref{sec:approach} describes our TagGNN approach. Section~\ref{sec:exp} reports on the experimental results. We review the related work in Section~\ref{sec:relatedwork} and finally conclude the paper in Section~\ref{sec:conclusion}.

\section{TagGNN Approach}\label{sec:approach}
In this section, we first introduce the motivation of our model design and present an overview of TagGNN. Then, we describe the details of our model, including three parts: TagGNN-IT, TagGNN-QI, and their integration TagGNN. Finally, we show the training and inference strategies for tag prediction. 

\subsection{Motivation and Overview}
\subsubsection{Motivation}
Nowadays, there is a trend to apply GNNs to enhancing text categorization tasks~\cite{GNN_text1,GNN_text2,GNN_text3}. Inspired by these studies, we explore the use of GNNs for item tagging in IR. Different from textual categorization, our work aims to address the following unique challenges.

Firstly, item tagging problems usually have a large tag space (more than thousands). It is desired to capture the rich semantic relationships among tags. Taking Figure~\ref{fig:app_demo} as an example, Pokemon has two strongly correlated tags, i.e., AR (Augmented Reality) and VR (Virtual Reality). Such tag correlations are indicative of the strong co-existence or non-existence for related tags. Existing GNN methods mostly model text categorization as a node classification problem, since the number of categories is usually small ($\sim$tens). This, however, ignores the dependency of category labels. 

Secondly, query information is readily available in IR tasks. While items lack concise textual descriptions, 
it is desired to join external information from query logs. For example, when a user search "chat" and download the app "Facebook", it potentially implies that the app is functionally related to "chat". Thus, tags like "chat" and "social" may be good candidates. The frequency of query-item interactions reveal the strength of such semantic correlations. How to effectively utilize the large amount of query information is an essential problem to build an accurate tagging system.  

Thirdly, while existing item nodes mostly have edge connections in the graph,  there are many new items everyday in the platform. These items have no links to either tag nodes or query nodes. This imposes a unique challenge for GNNs to deal with both full tag prediction and partial tag completion cases. 

\subsubsection{Overview}
\begin{figure*}[htbp]
    \begin{center}
    \includegraphics[width = 1\linewidth]{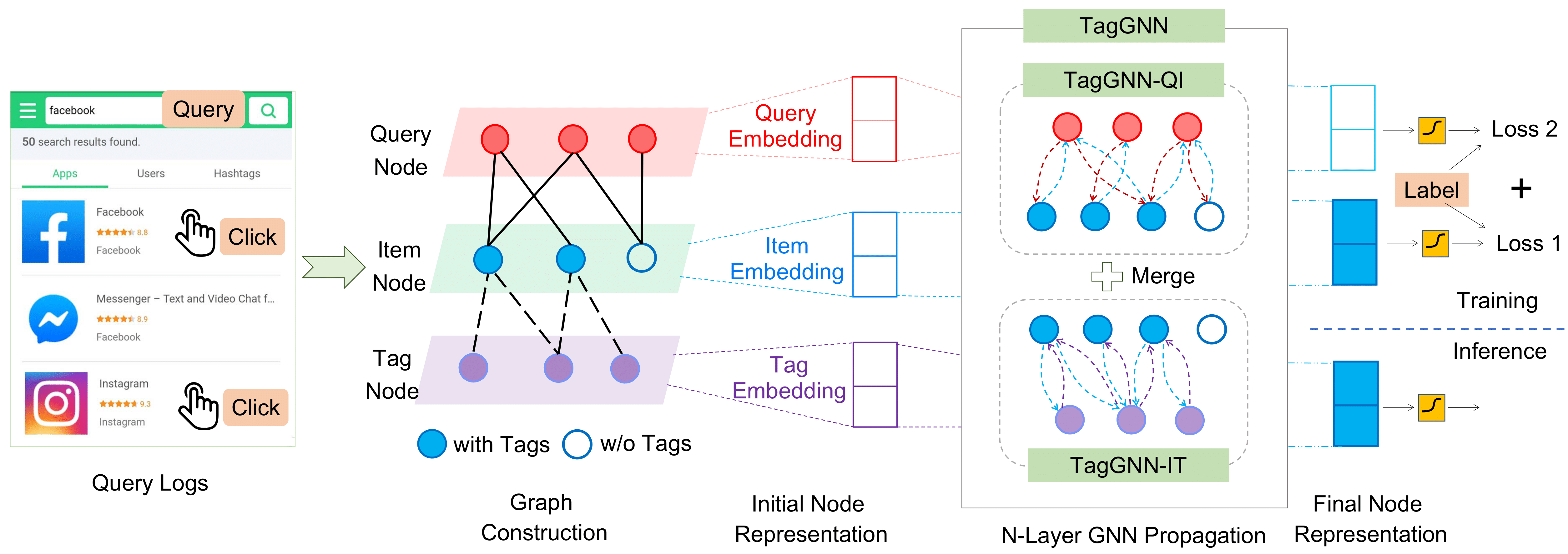}
    \end{center}
    \vspace{-2ex}
    \caption{Overview of TagGNN}
    \label{TagGNN}
\end{figure*}

To address the above three challenges, we present TagGNN, a GNN-based item tagging approach. Figure~\ref{TagGNN} provides an overview of TagGNN. Suppose that we have got the related queries of the items from the IR system, and we also know the items' corresponding tags. Then we build an undirected  tripartite graph to link query, item and tag together. The graph has three types of nodes, i.e., query, item and tag. Note that the item node can be unilateral or complete isolated if we do not know any related queries or existing tags (or both) of the item. Then, we employ TagGNN tailored for item tagging to propagate all of the information in the graph to get better item and tag representation. Finally, 
we compute the similarity between the item and  all tags and choose $K$ tags with the highest similarities as the our top$K$ prediction. 
In the following, we introduce TagGNN-IT, TagGNN-QI and TagGNN detail by detail.

\subsection{TagGNN-IT}
To fully exploit the interactions between items and tags, as well as correlations among tags, we treat the item tagging problem from the view of graph, modeling the multi-label classification as the link prediction problem in the graph. Specifically, we first build an undirected bipartite graph which has two types of nodes, i.e., item nodes and tag nodes. There will be an edge between the item node and the tag node if the item has the tag. Then TagGNN-IT learns new and powerful node representations in the graph for item tagging.

\subsubsection{Node Representation}
\label{initial representation}
We choose the item titles and the tag names as the initial features of item nodes and tag nodes respectively. Without loss of generality, suppose that the node contains a string of words $(w_1, w_2,..., w_n)$ extracted from content, title, name, description or others, we can use any models that can deal with the sequence, such as RNN and CNN, to get the initial representation $h$ of the node. 
In particular, since the tag sets are fixed, we add an extra id embedding, which is a one-hot vector, for each tag node. For simplicity, here we just average all of word embeddings as the initial node representation:  
\begin{equation}
h = \left\{
             \begin{array}{lr}
              \frac{1}{n}\sum_{i=1}^{n}{w_i},~~\text{if node is the item node}
             \\
              \frac{1}{n}\sum_{i=1}^{n}{w_i} + id,~~\text{if node is the tag node}&  
             \end{array}
\right.
\end{equation}
where $id$ is the one-hot id embedding for the tag node.

Note that traditional multi-label text classification approaches only use the id (one-hot) embeddings of tags, which does not explicitly consider the correlations between tags. We set the tags as nodes in the graph and fuse the semantic information of tags into the initial representations, not only better modeling the correlations between tags, but also improving the generalization ability of the model.

\subsubsection{TagGNN-IT Propagation}
\label{TagGNN-IT}
In the item-tag bipartite graph, the item node updates its representation by aggregating its neighbour tag nodes. Inspired by GAT~\cite{velivckovic2017graph}, we let the tag node first compute the similarity between every neighbour tag node with their semantic (node) representations in a common embedding space. Formally, for the item node $v$ and its neighbour tag node $w$, the similarity computation equation is:  
\begin{equation}
   \alpha_{vw}= \frac{\exp \left(\operatorname{LeakyReLU}\left(\mathbf{a}^{T}\left[{W} {h}_{v} \| {W} {h}_{w}\right]\right)\right)}{\sum_{k \in \mathcal{N}_{v}} \exp \left(\operatorname{LeakyReLU}\left(\mathbf{a}^{T}\left[{W} {h}_{v} \| {W} {h}_{k}\right]\right)\right)},
\end{equation}
where $W \in \mathbb{R}^{D}$ is a transformation matrix to transform both the item node and tag node into a common embedding space, $\mathbf{a}$ is a global context vector to determine the similarity between the two nodes. 

Then, based on their semantic similarity, the item node aggregates messages from all of neighbour tag nodes:
\begin{equation}
   {h}_{m}=\sigma(\sum_{w \in \mathcal{N}_{v}} \alpha_{v w} {W} {h}_{w}),
\end{equation}
where $h_m$ is the incoming aggregated message from neighbour tag nodes, $\sigma$ is the activation function (e.g., ReLU).

With the help of this attention mechanism, the item node can put more reasonable weights to its tag neighbours so that it can distinguish which tags are important, while which tags may be not informative and should be ignored. In this way, the item can more benefit from the representative tags and less affected by noisy tags.  

Finally, we fuse the message and the original item representation into a new item embedding space to get the new item representation. 
One such propagation is named one layer, and we stack $N$ such layers to capture higher-order neighbors' information. However, GNN often faces the over-smoothing problem as the number of layers gets deeper. To mitigate this and obtain more comprehensive representations, we adopt a gated skip-connection mechanism. The update equations are:
\begin{eqnarray}
    \hat{h}_{v} = \sigma(W_{item}(h_{v} + h_m)), \label{hete}\\
    z = \operatorname{sigmoid}(U_1\hat{h}_{v} + U_2h_{v} + b),\\
    h_{v}^{new} = z \odot \hat{h}_{v} + (1 - z) \odot h_v
\end{eqnarray}
where $W_{item}$ is the transformation matrix to the new \textbf{item} embedding space, $U_1$, $U_2$, $b$ are trainable parameters. $z$ controls the proportion of the original representation and the new representation to get the final new representation.

The tag nodes follow the similar propagation operations except that they have their own transformation matrix $W_{tag}$ when fusing the aggregated message from item nodes (equation \ref{hete}).

\subsubsection{Loss}
\label{itloss}
We deem the item tagging problem as a link prediction problem in the graph, i.e., to predict which tag nodes should be linked to the target item node, which can leverage enriched tag representations to improve the performance.Specifically, we compute dot-product similarity between the item representation $h_i$ and tag representation $h_t$, and compute its binary cross-entropy loss $\mathcal{L}_{LP}$ with the ground truth (0 or 1 represents link or not link):
\begin{equation}
    \mathcal{L}_{LP}(h_i, h_t, y) = BCE(y,~h_{i}\cdot h_{t})
\end{equation}
where $BCE$ is binary cross-entropy loss, $y$ is the ground truth for the edge between the item node and the tag node.

\subsection{TagGNN-QI}
To effectively leverage the query information which has not been exploited in previous literature, we design another model named TagGNN-QI also from the graph view.
Specifically, we build a query-item bipartite graph from the interactions of query logs and items. There will be an (weighted) edge between the query node and the item node if they are interacted. The query-item edge can represent different meanings depending on different real scenarios. For example, in the App Store scenario (app tagging), the query-item edge may represent the click or download behavior for the app under the query, and the edge weight can be the click times or downloads.

\subsubsection{Edge Representation}
In TagGNN-QI, both node features and edge features are used. Similar to TagGNN-IT, we use the query contents and item titles as the initial features of the query and item nodes, and also average all the word embeddings as the initial node representations.
Here we focus on edge features.
As the edge may contain useful information, we also encode the initial edge representation for TagGNN-QI. 
Specifically, if the edge originally has a feature vector, we just keep it. If the edge weight is a scalar, we can use the weight to enhance the message passing through this edge by simply multiply the message with the weight scalar. Besides, if the weight range is very large, we can use some feature scaling strategies like min-max normalization or standardization to rescale. We can also perform feature discretization, e.g., binning~\cite{zheng2018feature} , to get the initial edge representation.
If the edge not has weight, we just set all edge weights to 1.

\subsubsection{TagGNN-QI Propagation}
We change the similarity computation so as to utilize the information contained in the edge. Formally, if the edge representation $e_{vw}$ is a vector, the similarity is:
\begin{equation}
   \alpha_{vw}=\frac{\exp \left(\operatorname{LeakyReLU}\left(\mathbf{a}^{T}\left[{W} {h}_{v} \| {W} {h}_{w} \| {e}_{vw}\right]\right)\right)}{\sum_{k \in \mathcal{N}_{v}} \exp \left(\operatorname{LeakyReLU}\left(\mathbf{a}^{T}\left[{W} {h}_{v} \| {W} {h}_{k} \| {e}_{vk}\right]\right)\right)},
\end{equation}
While if the edge representation $e_{vw}$ is a scalar, the similarity is:
\begin{equation}
   \alpha_{vw}=e_{vw} \times \frac{\exp \left(\operatorname{LeakyReLU}\left(\mathbf{a}^{T}\left[{W} {h}_{v} \| {W} {h}_{w}\right]\right)\right)}{\sum_{k \in \mathcal{N}_{v}} \exp \left(\operatorname{LeakyReLU}\left(\mathbf{a}^{T}\left[{W} {h}_{v} \| {W} {h}_{k}\right]\right)\right)},
\end{equation}
The notations and other operations are consistent with TagGNN-IT described in \ref{TagGNN-IT}.

\subsubsection{Loss}
Since TagGNN-QI does not have tag nodes, we model it as a regular node classification form. We use a multi-layer perceptron (MLP) to transform the item representation $h_i$ to a $N$ dimensional vector $d$ (where $N$ is the number of tags) and we compute the mean binary cross-entropy loss $\mathcal{L}_{NC}$ with the ground truth:
\begin{eqnarray}
    &&
    d = W_{nc} h_{i} + q,\\
    &&
    \mathcal{L}_{NC}(h_i, y) = \frac{1}{N}\sum_{t = 1}^{N}{BCE}(d_t, y_t)
\end{eqnarray}
where $W_{nc}$ and $q$ are trainable parameters of MLP, $y_t$ is the ground truth between the item and the $t$-th tag, $d_t$ is the $t$-th dimension of $d$.

\subsection{TagGNN}
To solve all the three limitations simultaneously, we integrate TagGNN-IT and TagGNN-QI to a unified model named TagGNN, which inherits both their advantages. Specifically, we merge the former two bipartite graphs to one tripartite graph which has three types of nodes, i.e., query nodes, item nodes and tag nodes. The edges are the same as in the original graphs, and the initial representations of nodes and edges are also the same as before.
We perform message passing in this unified tripartite graph following the same propagation strategies described in TagGNN-IT and TagGNN-QI. Therefore, the item node will simultaneously get the messages from both query nodes and tag nodes to update its representation. TagGNN also deal with the item tagging as a link prediction problem, So, its loss form is same as TagGNN-IT described in \ref{itloss} (i.e., $\mathcal{L}_{LP}$).

\subsection{Training and Inference}
\label{training}
In this part, we introduce how the three models are trained and used for item tagging.
\subsubsection{Training}
In the graph, there may exist isolated test item nodes which we are unaware of any query or tag information about them, and we stipulate that their representations will not be updated by the GNN propagation. As the majority of training item nodes are not isolated, there will be a training-testing exposure bias, seriously reducing the the prediction precision of isolated test item nodes.

To empower the model with the ability to handle this ``cold start'' problem, we add a dual loss in addition to the primary loss during training. Specifically, the primary loss $\mathcal{L}_{1}$ is computed with the new learned item representation and the new learned tag representation. While the dual loss $\mathcal{L}_{2}$ is computed with the initial item representation and the new learned tag representation.
Finally we optimize the model by reducing these two losses together with stochastic gradient descent algorithms. Formally, for TagGNN-IT and TagGNN:
\begin{eqnarray}
    &&
    \mathcal{L}_{1} = \mathcal{L}_{LP}(h_{item}^{new}, h_{tag}^{new}, y), \\
    &&
    \mathcal{L}_{2} = \mathcal{L}_{LP}(h_{item}, h_{tag}^{new}, y),
\end{eqnarray}
for TagGNN-QI:
\begin{eqnarray}
    &&
    \mathcal{L}_{1} = \mathcal{L}_{NC}(h_{item}^{new}, y), \\
    &&
    \mathcal{L}_{2} = \mathcal{L}_{NC}(h_{item}, y),
\end{eqnarray}
The final optimization objective is:
\begin{eqnarray}
    &&
    \mathcal{L} = \mathcal{L}_{1} + \gamma \mathcal{L}_{2} 
\end{eqnarray}
where $\gamma$ is the hyper-parameter to adjust the proportion of $\mathcal{L}_{1}$ and $\mathcal{L}_{2}$. 

\subsubsection{Inference}
When inference, since the model has been optimized to be able to deal with isolated item nodes, we do not distinguish whether the item node is isolated or not. For TagGNN-IT and TagGNN, we compute similarities between the item representation (after propagation) and all of tags representations (after propagation), and choose $K$ tags with the highest similarities as the result.
For TagGNN-QI, we transform the item representation (after propagation) to a $N$ dimensional vector and choose the $K$ tags corresponding to the $K$ largest dimensions of the vector as the result.

\section{Experiment}\label{sec:exp}
In this section, we conduct experiments on two datasets about advertisement tagging and application tagging, aiming to answer the following questions:
\begin{itemize}
    \item \textbf{Q1}: How does TagGNN perform compared with the state-of-the-art item tagging related approaches on our tasks?
    \item \textbf{Q2}: Does the dual loss $\mathcal{L}_2$ really improve the performance of TagGNN? What is the impact of tag name embeddings? Does  heterogeneity of TagGNN take effect? 
    \item \textbf{Q3}: How do different designs (e.g., the number of TagGNN layers, the types of GNN) influence the performance of TagGNN?
    
\end{itemize}

\subsection{Dataset}
We perform experiments on the following two real-world datasets.

\textbf{KDDCup-2012}: This public dataset is originally provided by KDD Cup 2012 track2 competition for CTR prediction. Its training instances derived from session logs of the Tencent proprietary search engine, soso.com. From the dataset, we can get the \textbf{advertisements} (items),  \textbf{queries} that trigger the advertisements, and the \textbf{keywords} (tags) of the advertisements. We preprocess this dataset for advertisement tagging. Specifically, we process the dataset to satisfy the following three limitations:
\begin{itemize}
   \item In Query-Ad graph, every advertisement node link at least 20 query nodes, and every query node link at least 20 advertisement nodes.
   \item In Ad-Keyword graph, every advertisement node link at least 5 keyword nodes, and every keyword nodes link at least 15 advertisement nodes. 
   \item Each word in the vocabulary should appears at least 5 times.
\end{itemize}

\textbf{Huawei-Dataset}: It is a industrial dataset derived from a business company's App Store. To make it non-representative of the online app search traffic, we randomly sample a subset from original data, but cover both apps without tags and apps with partial tags, and use one week query logs. 
A query is related to the app when the user clicks or downloads the app searching with this query. The query-app edges have weights, which represents the  downloads of the app under the query. We use this dataset for app tagging.

The statistics of the two datasets are presented in Table\ref{dataset}.

\begin{table}[]
\centering
\caption{Dataset Statistics. "Avg. Queries" and "Avg. Tags" represent the average number of queries and tags associated to an item, respectively.}
\label{dataset}
\resizebox{0.47\textwidth}{!}{%
\begin{tabular}{ccccccc}
\hline
Dataset & \#Query & \#Item & \#Tag & \#Vocab & Avg. Queries & Avg. Tags \\ \hline
KDDCup-2012 & 92380 & 18861 & 9140 & 6620 & 89.4 & 13.8 \\
Huawei-Dataset & 47305 & 34166 & 2636 & 18601 & 5.8 & 3.6 \\ \hline
\end{tabular}%
}
\end{table}

\subsection{Experimental Setup}
We consider two types of tagging tasks. The first is \textbf{full tag prediction}, which means that we do not know any existing tags of the item and we should predict all of its tags. The second is \textbf{tag completion}, which means that we have known some tags of the item and we want to predict its remaining tags. For the second task, in our experiment, we randomly choose two tags of each item to predict and set its remaining tags as known tags.

For KDDCup-2012 dataset, we randomly choose 14861, 2000, 2000 advertisements for training, validation and test respectively.
In the validation and test parts, 1000 advertisements are used for full tag prediction and another 1000 advertisements are used for tag completion. For Huawei-Dataset, we randomly choose 28166, 3000, 3000 apps for training, validation and test respectively. Similarly, in the validation and test set, 1500 apps are for full tag prediction and another 1500 apps are for tag completion. 

The embedding size of the node and the word are both set to 200. The number of TagGNN layer is set to 2. TagGNN is trained with Adam optimizer, with 0.003 learning rate. Besides, we use standardization to normalize edge weights (we leave feature discretization in the future work).
We apply 0.5 feature dropout rate to alleviate overfitting. We stop training the model when the validation error plateaus. 
We use \textbf{Precision@K}, which is a common metric for multi-label classification task, as our evaluation metric.

\subsection{Baselines}
In order to verify the validity of TagGNN, we compare it with the following baselines\footnote{The embedding size is uniformly set to 200 for all baselines if not specified. Part of baselines are experimented with \url{https://github.com/Tencent/NeuralNLP-NeuralClassifier}.}:
\begin{itemize}
    \item \textbf{FastText-I}:  FastText~\cite{fasttext} is a simple and efficient text classification approach which averages the word/n-grams embeddings as the document embedding, then feeds the document embedding into a linear classifier. We use it to do multi-label text classification with item titles.
    \item \textbf{FastText-QI}: The only difference with FastText-I is that we concatenate the item title with its top-10 queries' contents as the new initial features.
    \item \textbf{Transformer-I}: This baseline follows the multi-label classification model~\cite{MTL_text4} that applies the most commonly used Transformers as the text encoder. We use item title as input. 
    \item \textbf{Transformer-QI}: The only difference with Transformer-I is that we concatenate the item titles with its top-10 queries' contents as the new initial features.
    \item \textbf{XmlCNN-QI}: 
    XmlCNN~\cite{MTL_text3} is a multi-label classification model that follows TextCNN~\cite{kim2014convolutional} to use CNNs as the text encoder. We use the concatenation of item title and query content as input. We set kernel sizes to \{2,3,4\} and use 100 kernels for each kernel size.
    \item \textbf{TextRNN-QI}: TextRNN~\cite{liu2016recurrent} is a  frequently-used text classification method which employ RNN with multi-task learning to encode the text. We use it to do multi-text classification with query contents and item titles. We use one-layer bidirectional RNN and the hidden embedding size is set to 200.
    \item \textbf{SimRank-QI}: SimRank~\cite{jeh2002simrank} is a popular graph-based approach that exploits the node-to-node relationships based on the topology of the graph. We propagate tags in the query-item bipartite graph based on SimRank to predict new tags for items.   
    \item \textbf{ML-GCN-I}: ML-GCN~\cite{image_tagging_1} is a recently published work that learns the label correlations via GCNs for image-based multi-label classification. We extend it to text-based classification and use FastText (performed best in experiments) as the textual encoder. We set $\tau$ to 0.1 and 0.3 for KDDCup-2012 and Huawei-Dataset respectively to build the needed label graph. Other settings are consistent with the original paper.
    \item \textbf{ML-GCN-QI}: The only difference with ML-GCN-I is that we change the main model to \textbf{TagGNN-QI}.
\end{itemize}

\begin{table*}[]
\centering
\caption{Performance comparison of different models. ``Without Tags'' indicates that items have no tags before prediction, and ``Partial Tags'' means that items have incomplete tags and need tag completion. TagGNN-IT, TagGNN-QI and TagGNN show our approaches.}
\label{result}
\resizebox{\textwidth}{!}{%
\begin{tabular}{c|c|c|ccc|ccc|ccc|ccc}
\hline
 &  &  & \multicolumn{6}{c|}{KDDCup-2012} & \multicolumn{6}{c}{Huawei-Dataset} \\ \cline{4-15} 
 & Model & Features & \multicolumn{3}{c|}{Without Tags} & \multicolumn{3}{c|}{Partial Tags} & \multicolumn{3}{c|}{Without Tags} & \multicolumn{3}{c}{Partial Tags} \\ \cline{4-15} 
 &  &  & P@1 & P@3 & P@5 & P@1 & P@3 & P@5 & P@1 & P@3 & P@5 & P@1 & P@3 & P@5 \\ \hline
 & FastText-I & Item Text & 0.405 & 0.352 & 0.331 & 0.158 & 0.134 & 0.105 & 0.529 & 0.386 & 0.286 & 0.392 & 0.265 & 0.197 \\
 & FastText-QI & Item \& Query Text & 0.581 & 0.510 & 0.470 & 0.286 & 0.190 & 0.143 & 0.688 & 0.492 & 0.354 & 0.515 & 0.340 & 0.246 \\
Text based & Transformer-I & Item Text & 0.373 & 0.332 & 0.311 & 0.175 & 0.124 & 0.098 & 0.471 & 0.338 & 0.249 & 0.358 & 0.244 & 0.185 \\
 & Transformer-QI & Item \& Query Text & 0.443 & 0.393 & 0.363 & 0.161 & 0.124 & 0.097 & 0.608 & 0.432 & 0.313 & 0.477 & 0.314 & 0.227 \\
 & XmlCNN-QI & Item \& Query Text & 0.371 & 0.327 & 0.302 & 0.112 & 0.083 & 0.064 & 0.515 & 0.356 & 0.259 & 0.341 & 0.225 & 0.163 \\
 & TextRNN-QI & Item \& Query Text & 0.484 & 0.424 & 0.387 & 0.169 & 0.117 & 0.089 & 0.615 & 0.428 & 0.308 & 0.379 & 0.255 & 0.186 \\ \hline
Graph based & SimRank-QI & Query-Item Graph & 0.559 & 0.510 & 0.479 & 0.171 & 0.144 & 0.125 & 0.577 & 0.421 & 0.299 & 0.499 & 0.342 & 0.243 \\ \hline
 & ML-GCN-I & Tag-Tag Graph & 0.365 & 0.311 & 0.296 & 0.191 & 0.148 & 0.113 & 0.414 & 0.342 & 0.251 & 0.414 & 0.276 & 0.200 \\
 & ML-GCN-QI & Query-Item \& Tag-Tag Graph & 0.742 & 0.672 & 0.625 & 0.385 & 0.273 & 0.193 & 0.721 & 0.519 & 0.371 & 0.612 & 0.388 & 0.272 \\ \cline{2-15} 
Graph \& Text based & TagGNN-IT & Item-Tag Graph & 0.438 & 0.326 & 0.280 & 0.342 & 0.250 & 0.187 & 0.539 & 0.362 & 0.264 & 0.444 & 0.286 & 0.209 \\
 & TagGNN-QI & Query-Item Graph & 0.755 & 0.688 & 0.643 & 0.403 & 0.295 & 0.214 & 0.730 & 0.520 & 0.379 & 0.618 & 0.395 & 0.276 \\
 & TagGNN & Query-Item-Tag Graph & \textbf{0.823} & \textbf{0.741} & \textbf{0.683} & \textbf{0.449} & \textbf{0.330} & \textbf{0.236} & \textbf{0.743} & \textbf{0.534} & \textbf{0.381} & \textbf{0.644} & \textbf{0.416} & \textbf{0.288} \\ \hline
\end{tabular}%
}
\end{table*}

\subsection{Performance Comparison (Q1)}
\label{performance}
The comparative results are summarized in Table~\ref{result}. In the following, we discuss the results of two tasks, i.e., full tag prediction and tag completion respectively.
\subsubsection{Results of Full Tag Prediction}
We have the following observations about the results of full tag prediction task:
\begin{itemize}
    \item Our final TagGNN substantially outperforms all the other baselines on both two datasets, verifying the effectiveness of our model to solve the full tag prediction task. In particular, TagGNN improves the strongest baseline TagGNN-QI (also ours) by 6.8\% and 5.3\% in P@1 and P@5 on KDDCup-2012 dataset.
    We attribute such notable improvements to the novel and powerful design of TagGNN that can benefit from both explicit and implicit interactions and representation fusions among queries, items and tags.
    \item Query information is very useful and can be easily utilized to solve the full tag prediction task. It is obvious that the precision gains a huge improvement (7\% to 38.5\% for KDDCup-2012, and 13.7\% to 30.7\% for Huawei-Dataset) for all baselines after fusing the query information, which strongly proves the importance of queries. Note that the gains of ML-GNN-QI compared with ML-GNN-I are also mostly originated from TagGNN-QI since their major difference lies in the main models of ML-GNN. Thus, comparatively speaking, TagGNNs get the biggest percentages of boost from the query information, demonstrating that TagGNNs can utilize queries better than other baselines.
    \item Graph-based SimRank-QI and ML-GCN-I are inferior to text-based FastText-QI and Transformer-I respectively. This phenomenon shows that not all graph-based or graph\&text-based methods can beat the traditional text-based methods. How to use all the information in the form of graph is the real key for graph based methods, not the graph form itself. This also proves that TagGNN can take advantage of graph information more effectively. 
    \item Unexpectedly, ML-GCN-QI performs worse than TagGNN-QI. Since the main model of ML-GCN-QI we used is just TagGNN-QI, it demonstrates that the label (tag) embedding strategy proposed in ML-GCN is not effective on item tagging task, and further proves the effectiveness of the way that TagGNN leveraging the tags. 
\end{itemize}

\subsubsection{Results of Tag Completion} 
We have the following observations about the results of tag completion task:
\begin{itemize}
    \item TagGNN still achieves the best performance across the two datasets on tag completion task, demonstrating the comprehensive superiority of TagGNN than other baselines. Specifically, it can beat the strongest baseline TagGNN-QI (also ours) by 4.6\% and 2.6\% in P@1 on KDDCup-2012 dataset and Huawei-Dataset, and hugely surpasses all the text-based approaches,  which is a relative good performance.
    \item When queries are not available , TagGNN-IT outperforms all the other baselines, i.e., FastText-I, Transformer-I and ML-GCN-I on the two datasets. More notably, on KDDCup-2012 dataset, even if the other baselines using queries, TagGNN-IT can still outperforms them in most cases. It may be because that the number of tags in KDDCup-2012 is larger than Huawei-Dataset (as shown in Tabel~\ref{dataset}), which boosts the TagGNN to better release its potency. Such an excellent performance of TagGNN-IT also verifies that the design of our TagGNN framework has strong ability to leverage existing tags so as to improve the performance of tag completion task. 
    
\end{itemize}

In addition, some readers may wonder why the results of full tag prediction seem to be better than the results of tag completion as shown in Table~\ref{result}? Here is an illustration:

These two tasks have different numbers of ground truth tags. Note that there are only 2 ground truth tags for tag completion task. So, compared with the larger ground truth set of full tag prediction task, it is more difficult to hit the ground truth tags in tag completion task, leading to an illusion that the tag completion's precision is lower than the full tag prediction's. 

\begin{figure}[htbp]
    \centering  
    \subfigure[KDDCup-2012]{

    \includegraphics[width=0.48\columnwidth]{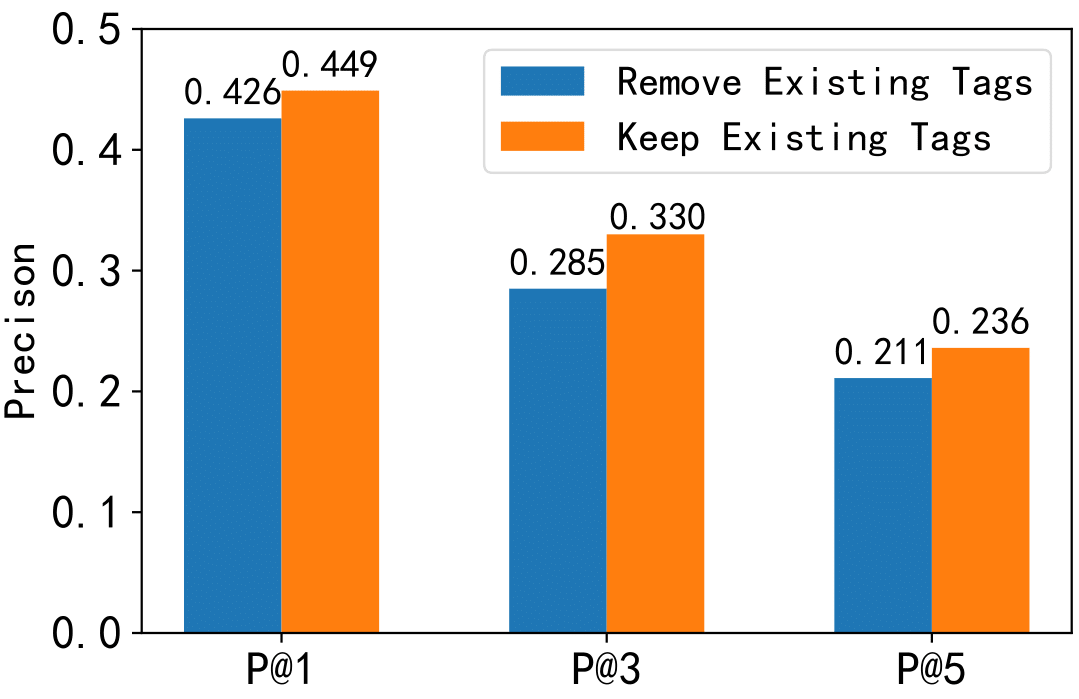}}
    \subfigure[Huawei-Dataset]{

    \includegraphics[width=0.48\columnwidth]{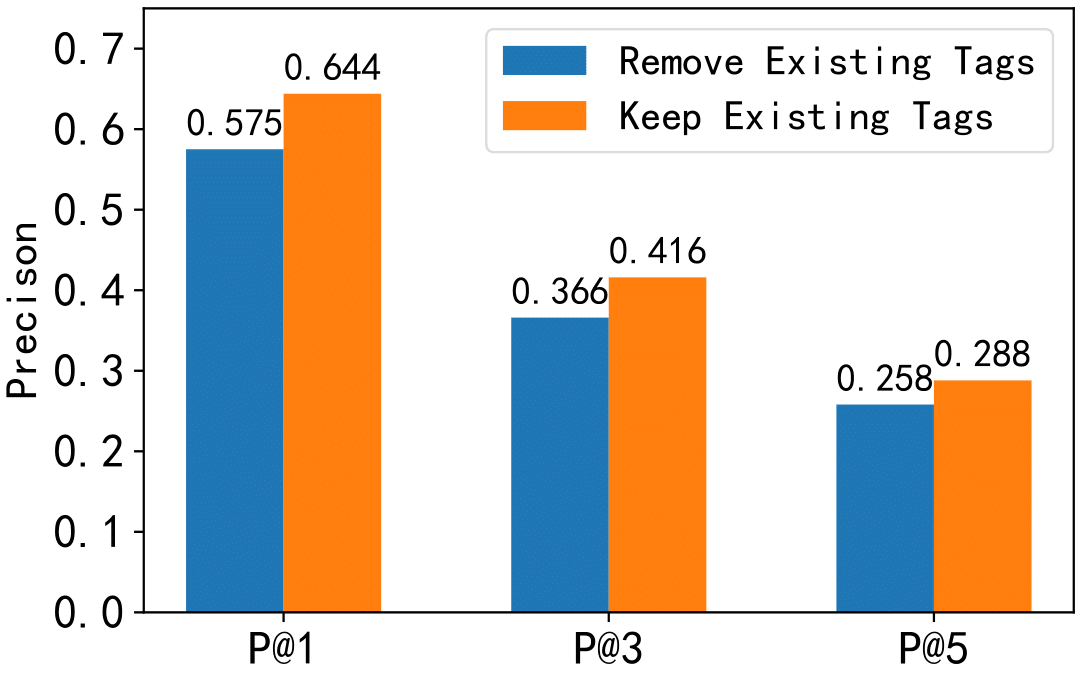}}
    
    \caption{Results of removing/keeping existing tags for the tag completion task. }
    \label{illus}
  \end{figure}

To further demonstrate that TagGNN can really leverage the existing tags to improve the performance of tag completion, we remove all the existing tags of items in the test set, and retrain the model to test its performance. Results are shown in Figure~\ref{illus}. It is obvious that the performance gets worse after removing the existing tags, which verifies our illustration.

\begin{table*}[]
\centering
\small
\caption{Performance comparison of ablation study. ``TagGNN w/o $\mathcal{L}_2$ \& TNE'' represents TagGNN without both $\mathcal{L}_2$ and tag name embeddings,  ``TagGNN w/o $\mathcal{L}_2$'' represents TagGNN only without $\mathcal{L}_2$, ``TagGNN-homogeneous'' represents TagGNN using homogeneous update function. Similar notations for others.}
\label{study}
\resizebox{0.85\textwidth}{!}{%
\begin{tabular}{c|ccc|ccc|cccccc}
\hline
\multirow{3}{*}{Model} & \multicolumn{6}{c|}{KDDCup-2012} & \multicolumn{6}{c}{Huawei-Dataset} \\ \cline{2-13} 
 & \multicolumn{3}{c|}{Without Tags} & \multicolumn{3}{c|}{Partial Tags} & \multicolumn{3}{c|}{Without Tags} & \multicolumn{3}{c}{Partial Tags} \\ \cline{2-13} 
 & P@1 & P@3 & P@5 & P@1 & P@3 & P@5 & P@1 & P@3 & \multicolumn{1}{c|}{P@5} & P@1 & P@3 & P@5 \\ \hline\hline
TagGNN-IT & 0.438 & \textbf{0.326} & \textbf{0.280} & 0.342 & 0.250 & 0.187 & \textbf{0.539} & \textbf{0.362} & \multicolumn{1}{c|}{\textbf{0.264}} & 0.444 & 0.286 & 0.209 \\\hline
TagGNN-IT w/o $\mathcal{L}_2$ & 0.016 & 0.011 & 0.010 & \textbf{0.360} & 0.274 & \textbf{0.200} & 0.291 & 0.199 & \multicolumn{1}{c|}{0.149} & \textbf{0.447} & \textbf{0.294} & 0.207 \\
TagGNN-IT w/o $\mathcal{L}_2$ \& TNE & 0.012 & 0.010 & 0.009 & 0.332 & \textbf{0.258} & 0.192 & 0.289 & 0.200 & \multicolumn{1}{c|}{0.148} & \textbf{0.447} & 0.293 & \textbf{0.210} \\
TagGNN-IT-homogeneous & \textbf{0.439} & 0.320 & 0.274 & 0.333 & 0.241 & 0.183 & 0.529 & 0.352 & \multicolumn{1}{c|}{0.261} & 0.428 & 0.277 & 0.206 \\\hline\hline
TagGNN-QI & \textbf{0.755} & \textbf{0.688} & \textbf{0.643} & \textbf{0.403} & \textbf{0.295} & \textbf{0.214} & \textbf{0.723} & 0.520 & \multicolumn{1}{c|}{\textbf{0.379}} & \textbf{0.618} & \textbf{0.395} & \textbf{0.276} \\ \hline
TagGNN-QI w/o $\mathcal{L}_2$ & 0.746 & 0.679 & 0.639 & 0.402 & 0.288 & 0.211 & \textbf{0.723} & 0.517 & \multicolumn{1}{c|}{0.378} & 0.611 & 0.393 & \textbf{0.276} \\
TagGNN-QI w/o $\mathcal{L}_2$ \& TNE & 0.747 & 0.675 & 0.635 & 0.401 & 0.286 & 0.209 & 0.715 & 0.516 & \multicolumn{1}{c|}{0.371} & 0.606 & 0.392 & 0.272 \\
TagGNN-QI-homogeneous & 0.745 & 0.675 & 0.632 & 0.391 & 0.280 & 0.207 & 0.722 & \textbf{0.524} & \multicolumn{1}{c|}{0.369} & 0.608 & 0.381 & 0.268 \\\hline\hline

TagGNN & \textbf{0.823} & \textbf{0.741} & \textbf{0.683} & \textbf{0.449} & \textbf{0.330} & \textbf{0.236} & \textbf{0.743} & \textbf{0.534} & \multicolumn{1}{c|}{\textbf{0.381}} & \textbf{0.644} & \textbf{0.416} & \textbf{0.288} \\ \hline
TagGNN w/o $\mathcal{L}_2$ & 0.791 & 0.719 & 0.664 & 0.442 & 0.316 & 0.235 & 0.721 & 0.521 & \multicolumn{1}{c|}{0.377} & 0.637 & 0.409 & 0.282 \\
TagGNN w/o $\mathcal{L}_2$ \& TNE & 0.789 & 0.715 & 0.661 & 0.426 & 0.306 & 0.227 & 0.711 & 0.519 & \multicolumn{1}{c|}{0.361} & 0.624 & 0.391 & 0.275 \\
TagGNN-homogeneous & 0.807 & 0.724 & 0.673 & 0.417 & 0.315 & 0.227 & 0.732 & 0.527 & \multicolumn{1}{c|}{0.375} & 0.629 & 0.404 & 0.282 \\\hline\hline

\end{tabular}%
}
\end{table*}

\subsection{Ablation Study (Q2)}
In this section, we study how three particular components of TagGNN, i.e., $\mathcal{L}_2$ (see ~\ref{training}), tag name embeddings (see ~\ref{initial representation}) and heterogeneity affect the performance and answer \textbf{Q2}.

\begin{table*}[!t]
\centering
\caption{Performance comparison of TagGNN with different number of propagation layers.}
\label{ab-layer}
\resizebox{0.73\textwidth}{!}{%
\begin{tabular}{c|ccc|ccc|cccccc}
\hline
\multirow{3}{*}{Model} & \multicolumn{6}{c|}{KDDCup-2012} & \multicolumn{6}{c}{Huawei-Dataset} \\ \cline{2-13} 
 & \multicolumn{3}{c|}{Without Tags} & \multicolumn{3}{c|}{Partial Tags} & \multicolumn{3}{c|}{Without Tags} & \multicolumn{3}{c}{Partial Tags} \\ \cline{2-13} 
 & P@1 & P@3 & P@5 & P@1 & P@3 & P@5 & P@1 & P@3 & \multicolumn{1}{c|}{P@5} & P@1 & P@3 & P@5 \\ \hline
TagGNN-1 & 0.786 & 0.697 & 0.649 & 0.412 & 0.302 & 0.218 & 0.676 & 0.474 & \multicolumn{1}{c|}{0.338} & 0.553 & 0.351 & 0.245 \\
TagGNN-2 & \textbf{0.823} & \textbf{0.741} & \textbf{0.683} & \textbf{0.449} & \textbf{0.330} & 0.236 & \textbf{0.743} & \textbf{0.534} & \multicolumn{1}{c|}{\textbf{0.381}} & \textbf{0.644} & \textbf{0.416} & \textbf{0.288} \\
TagGNN-3 & 0.815 & 0.735 & 0.674 & 0.432 & 0.321 & \textbf{0.238} & 0.732 & 0.515 & \multicolumn{1}{c|}{0.364} & 0.641 & 0.411 & 0.284 \\
TagGNN-4 & 0.811 & 0.728 & 0.670 & 0.428 & 0.315 & 0.234 & 0.728 & 0.507 & \multicolumn{1}{c|}{0.361} & 0.633 & 0.409 & 0.283 \\ \hline
\end{tabular}%
}
\end{table*}

\begin{table*}[!t]
\centering
\caption{Performance comparison of different types of GNN.}
\label{ab-GNNtypes}
\resizebox{0.73\textwidth}{!}{%
\begin{tabular}{c|ccc|ccc|cccccc}
\hline
\multirow{3}{*}{Model} & \multicolumn{6}{c|}{KDDCup-2012} & \multicolumn{6}{c}{Huawei-Dataset} \\ \cline{2-13} 
 & \multicolumn{3}{c|}{Without Tags} & \multicolumn{3}{c|}{Partial Tags} & \multicolumn{3}{c|}{Without Tags} & \multicolumn{3}{c}{Partial Tags} \\ \cline{2-13} 
 & P@1 & P@3 & P@5 & P@1 & P@3 & P@5 & P@1 & P@3 & \multicolumn{1}{c|}{P@5} & P@1 & P@3 & P@5 \\ \hline
GCN & 0.771 & 0.683 & 0.635 & 0.441 & 0.313 & 0.227 & 0.717 & 0.507 & \multicolumn{1}{c|}{0.359} & 0.594 & 0.386 & 0.269 \\
GraphSAGE & 0.793 & 0.712 & 0.663 & 0.441 & 0.305 & 0.221 & 0.675 & 0.465 & \multicolumn{1}{c|}{0.331} & 0.592 & 0.372 & 0.258 \\
GAT & 0.806 & 0.725 & 0.671 & 0.424 & 0.306 & 0.223 & 0.722 & 0.515 & \multicolumn{1}{c|}{0.364} & 0.623 & 0.394 & 0.273 \\\hline
TagGNN & \textbf{0.823} & \textbf{0.741} & \textbf{0.683} & \textbf{0.449} & \textbf{0.330} & \textbf{0.236} & \textbf{0.743} & \textbf{0.534} & \multicolumn{1}{c|}{\textbf{0.381}} & \textbf{0.644} & \textbf{0.416} & \textbf{0.288} \\ \hline
\end{tabular}%
}
\end{table*}

\subsubsection{Dual Loss $\mathcal{L}_2$}
As described in~\ref{training}, we add the dual loss $\mathcal{L}_2$ to deal with the cold start problem of isolated items. Here, we perform a ablation study to verify the validity of this strategy by removing $\mathcal{L}_2$ when training.
Table~\ref{study} shows the experimental results of TagGNNs training with $\mathcal{L}_2$ and without $\mathcal{L}_2$, and we have the following findings:
\begin{itemize}
    \item Without $\mathcal{L}_2$, performances of TagGNN-IT for full tag prediction on both two datasets drop sharply, indicating that the model nearly loses the ability to handle the cold start problem.
    Such performance degradation is due to the fact that When the graph has no query nodes, all training item nodes still have neighbour tag nodes. However, the test item nodes for full tag prediction will have no neighbours to aggregate, leading to a huge gap between training and inference.
    
    \item Although $\mathcal{L}_2$ may slightly hurt the precision of TagGNN-IT on tag completion task, it is trivial compared with the huge improvement on full tag prediction task. On the whole, our proposed dual loss $\mathcal{L}_2$ is really an effective way to handle the isolated nodes.
    
    \item Queries are quite informative and powerful to greatly alleviate the gap mentioned above. From the results of ``$\operatorname{TagGNN-QI}$ $w/o \ \mathcal{L}_{2}$''  and ``$\operatorname{TagGNN} \ w/o \  \mathcal{L}_{2}$'', we find that when the query information is available, $\mathcal{L}_2$ may be cannot bring very notable promotion as before. But it is still a valid auxiliary to improve accuracy.
    \item Moreover, jointly considered Table~\ref{result} and Table~\ref{study}, we find that even without query information and $\mathcal{L}_2$, ``$\operatorname{TagGNN-IT} w/o$ $\mathcal{L}_2$'' is still much better than traditional multi-label text classification methods on tag completion task. This demonstrates that the mode of TagGNN-IT can more effectively utilize the existing tag information to help solve tag completion task.
    \item We note that the heterogeneous GNN based HGAT model~\cite{GNN_text1} cannot be directly applied to item tagging. But considering that it models text categorization as node classification, we can take "{TagGNN-QI} w/o $\mathcal{L}_2$" as the approximate implementation of HGAT on item tagging. The results show that TagGNN is much better than HGAT on both datasets. 
\end{itemize}

\subsubsection{Tag Name Emeddings}
We introduced in~\ref{initial representation} that the initial representation of the tag node is the combination of its tag name embedding and tag id embedding. Note that the tag id embedding is always available since it is a one-hot embedding which is only related to the total number of tags. However, the tag name may be not visible during training the model in some situations (e.g., the company outsources the project of item tagging to others but it do not want to disclose the exact names of tags). Thus, here we remove the tag name embeddings from the initial node representation and see how it influence the performance of TagGNN. The experimental results are reported in Table~\ref{study}.

On the whole, the results show that the tag name embeddings (TNE) just bring slight improvement. But we believe the potential of TNE is much more than that. we are also considering how to better use the semantic information of tag names, such as exploring more fine-grained word-level interactions among items and tags. We leave it in our future work.

\subsubsection{ homogeneity and heterogeneity of TagGNN}
Considering that the query-item-tag tripartite graph is heterogeneous, we also design TagGNN to be heterogeneous, as embodied in its update function (equation \ref{hete}). To demonstrate the effectiveness of this heterogeneous design, we change the update function to be homogeneous, i.e., not distinguish the node types, and test the performance of TagGNN-QI, TagGNN-IT and TagGNN. We report the experimental results in Table~\ref{study}.

From the results, we can see that heterogeneous TagGNNs are nearly consistently better than homogeneous ones, demonstrating that setting different transformation matrices for different types of nodes is reasonable and valid, which can bring steady improvement.

\subsection{Design Choices of TagGNN (Q3)}
In this part, we research how different designs influence the performance from two perspectives.

\subsubsection{\textbf{Effect of Layer Numbers}}
To explore how the number of propagation layers affects the performance, we vary the number of model layers. Specially, we conduct experiments with the layer numbers in range of \{1, 2, 3, 4\}. Table~\ref{ab-layer} summarizes the experimental results, wherein TagGNN-X indicates the model with X layers. From the results, we have the following observations:

\begin{itemize}
    \item TagGNN-1 is obviously worse than TagGNN-2,3,4, indicating that only one propagation layer is not enough to reach an excellent performance. It is reasonable since one-layer GNN propagation can only capture the first-order neighbors' information. Hence, semantic relationships between query and query, item and item, tag and tag are not explicitly used, resulting in unsatisfactory performance. So it is necessary to stack at least two propagation layers.
    \item Stacking too much (larger than 3) layers will not bring additional promotion. Compared with TagGNN-2, only TagGNN-3 got a little gain (0.2\%) in P@5 (Paritial Tags of KDDCup-2012), verifying that two layers are enough for TagGNN. Too many layers may lead to redundancy that hurts performance.
\end{itemize}

\subsubsection{\textbf{Effect of Types of GNN}}
To verify the superiority of the propagation design of TagGNN, we replace the TagGNN with some other popular GNN models, e.g., GCN, GraphSAGE and GAT. For GraphSAGE, we choose its ``mean'' strategy. All corresponding settings are consistent with TagGNN. We show the experimental results in Table~\ref{ab-GNNtypes}.

The results shows that our TagGNN is clearly superior to all these representative GNNs. Specifically, GCN and GraphSAGE beat each other on two datasets but are worse than GAT. As for GAT, it is modestly inferior to TagGNN. It may be because that TagGNN can leverage additional edge information and has better representation fusion between two layers, which makes TagGNN more effective for item tagging.

\subsection{Expert Evaluation}
Before deploying the tagging model for production use, we need to perform a manual A/B testing by our operation team. Specifically, we randomly sample 540 apps from the test set of Huawei-Dataset, half for full tag prediction and half for tag completion. The sampling is performed uniformly to keep the proportion of each app category (e.g., game, study) consistent with the whole app corpus. In the full tag prediction setting, we predict top-5 tags for each app, while in the tag completion setting, we predict top-k tags to assure that each item has at least five tags. For example, if an item has 3 existing tags (3.6 on average), we set k$=$2. We generate two groups of app-tag samples predicted using both TagGNN and our production baseline model. This leads to a total of 4050 app-tag pairs. We randomly split the test samples and distribute them to four domain experts from our operation team. They assess the test samples one by one to check whether a tag is appropriate for an app. Finally, the expert evaluation results show that TagGNN achieves 81.1\% accuracy for full tag prediction, and 88.2\% accuracy for tag completion. Meanwhile, TagGNN achieves a 22.8\% relative improvement over the production baseline. The improvement is significant for production deployment.
\section{Related Work}\label{sec:relatedwork}
\subsection{Multi-Label Classification}
Multi-label classification~\cite{MTL_survey} is a widely-studied research topic, spanning multiple tasks such as text tagging~\cite{MTL_text3, MTL_text4} and image annotation~\cite{image_tagging_1}. Recent research efforts have been devoted to optimizing the content representation learning or exploring label dependencies for improvement. More specifically, Liu et al.~\cite{MTL_text3} and Chang et al.~\cite{MTL_text4} study the application of CNNs and transformers to enhance text-based multi-label classification, respectively. Chen et al.~\cite{image_tagging_1} investigate the use of GNNs to capture correlations among labels. All these studies assume rich contents. In contrast, we have to leverage external information (e.g., query logs) to enrich items. We also empirically compare TagGNN with them in Table~\ref{result}.

\subsection{Graph Neural Networks}
Graph neural networks (GNNs)~\cite{GNN_survey} has become a trending research topic. The research of GNNs successfully extends traditional convolutional neural networks to graph-structured data, leading to abundant applications such as text categorization~\cite{GNN_text2},  recommendation~\cite{Pinsage}, and link prediction~\cite{GraphSAGE}. Our work is inspired by these successful studies, and has been extended for item tagging. We empirically compare TagGNN with three representative GNN models, i.e., GCN~\cite{GCN}, GraphSAGE~\cite{GraphSAGE}, and GAT~\cite{velivckovic2017graph}. 

\subsection{GNN-based Text Categorization}
As a promising technique, GNNs have been recently adopted to boost text categorization tasks. In particular, Yao et al.~\cite{GNN_text2} propose the first use of graph convolution networks for text classification. But this work models each document as a graph node and cannot handle new documents that are not present in the graph during training. Later work~\cite{GNN_text1,GNN_text3} makes some extensions to tackle this issue. Especially, Hu et al.~\cite{GNN_text1} construct a topic-document-entity graph and model it using heterogeneous GNNs. This work is mostly closest to ours. However, the differences lie in that: 1) We model item tagging as a link prediction problem, instead of the node classification formulation in~\cite{GNN_text1}, which enables both full tag prediction and tag completion. 2) Our query and tag nodes, which naturally exist in IR tasks, provide multi-source information to enrich item representation, but topic and entity nodes are all intermediate information extracted from documents using preprocessing tools.


\section{Conclusion}\label{sec:conclusion}
In this paper, we present TagGNN, a tripartite graph neural network model for item tagging. Our model builds on the heterogeneous GNN techniques, but differs from other previous studies in three unique aspects: 1) Instead of node classification, TagGNN formulates item tagging as a novel link prediction problem. 2) TagGNN leverages query logs to enrich item representation and forms a query-item-tag tripartite graph that is unique for IR. 3) TagGNN is capable of making both full tag prediction and partial tag completion in a unified way. Experimental results on two large datasets validate the superiority of our TagGNN approach over existing methods. In addition, we perform an expert evaluation from our operation team and obtain quite positive results for production use.


\section{Acknowledgments}
This work is supported in part by the National Natural Science Foundation of Guangdong Province (2018A030313422), the National Natural Science Foundation of China (61972219, 61773229), the National Key Research and Development Program of China (2018YFB1800600, 2018YFB1800204), the R\&D Program of Shenzhen (JCYJ20190813174403598, JCYJ20190813165003837), and Overseas Cooperation Research Fund of Graduate School at Shenzhen, Tsinghua University (HW2018002), and the research fund of PCL Future Regional Network Facilities for Large-scale Experiments and Applications (PCL2018KP001).

\bibliographystyle{ACM-Reference-Format}
\balance
\bibliography{sigir}


\begin{thebibliography}{35}


\ifx \showCODEN    \undefined \def \showCODEN     #1{\unskip}     \fi
\ifx \showDOI      \undefined \def \showDOI       #1{#1}\fi
\ifx \showISBNx    \undefined \def \showISBNx     #1{\unskip}     \fi
\ifx \showISBNxiii \undefined \def \showISBNxiii  #1{\unskip}     \fi
\ifx \showISSN     \undefined \def \showISSN      #1{\unskip}     \fi
\ifx \showLCCN     \undefined \def \showLCCN      #1{\unskip}     \fi
\ifx \shownote     \undefined \def \shownote      #1{#1}          \fi
\ifx \showarticletitle \undefined \def \showarticletitle #1{#1}   \fi
\ifx \showURL      \undefined \def \showURL       {\relax}        \fi
\providecommand\bibfield[2]{#2}
\providecommand\bibinfo[2]{#2}
\providecommand\natexlab[1]{#1}
\providecommand\showeprint[2][]{arXiv:#2}

\bibitem[\protect\citeauthoryear{??}{man}{2018}]%
        {manual_tagging}
 \bibinfo{year}{2018}\natexlab{}.
\newblock \bibinfo{title}{Metadata and the Tagging Process at The New York
  Times}.
\newblock
\newblock
\urldef\tempurl%
\url{https://iptc.org/news/metadata-and-the-tagging-process-at-the-new-york-times}
\showURL{%
\tempurl}


\bibitem[\protect\citeauthoryear{Chalkidis, Fergadiotis, Malakasiotis, and
  Androutsopoulos}{Chalkidis et~al\mbox{.}}{2019}]%
        {MTL_text1}
\bibfield{author}{\bibinfo{person}{Ilias Chalkidis}, \bibinfo{person}{Manos
  Fergadiotis}, \bibinfo{person}{Prodromos Malakasiotis}, {and}
  \bibinfo{person}{Ion Androutsopoulos}.} \bibinfo{year}{2019}\natexlab{}.
\newblock \showarticletitle{Large-Scale Multi-Label Text Classification on {EU}
  Legislation}. In \bibinfo{booktitle}{\emph{Proceedings of the 57th Conference
  of the Association for Computational Linguistics ({ACL})}}.
  \bibinfo{pages}{6314--6322}.
\newblock


\bibitem[\protect\citeauthoryear{Chen, Hoi, Li, and Xiao}{Chen
  et~al\mbox{.}}{2016}]%
        {app_tagging}
\bibfield{author}{\bibinfo{person}{Ning Chen}, \bibinfo{person}{Steven C.~H.
  Hoi}, \bibinfo{person}{Shaohua Li}, {and} \bibinfo{person}{Xiaokui Xiao}.}
  \bibinfo{year}{2016}\natexlab{}.
\newblock \showarticletitle{Mobile App Tagging}. In
  \bibinfo{booktitle}{\emph{Proceedings of the Ninth {ACM} International
  Conference on Web Search and Data Mining (WSDM)}}. \bibinfo{pages}{63--72}.
\newblock


\bibitem[\protect\citeauthoryear{Chen, Wei, Wang, and Guo}{Chen
  et~al\mbox{.}}{2019}]%
        {image_tagging_1}
\bibfield{author}{\bibinfo{person}{Zhao{-}Min Chen},
  \bibinfo{person}{Xiu{-}Shen Wei}, \bibinfo{person}{Peng Wang}, {and}
  \bibinfo{person}{Yanwen Guo}.} \bibinfo{year}{2019}\natexlab{}.
\newblock \showarticletitle{Multi-Label Image Recognition With Graph
  Convolutional Networks}. In \bibinfo{booktitle}{\emph{{IEEE} Conference on
  Computer Vision and Pattern Recognition ({CVPR})}}.
  \bibinfo{pages}{5177--5186}.
\newblock


\bibitem[\protect\citeauthoryear{Florescu and Caragea}{Florescu and
  Caragea}{2017}]%
        {PositionRank}
\bibfield{author}{\bibinfo{person}{Corina Florescu} {and}
  \bibinfo{person}{Cornelia Caragea}.} \bibinfo{year}{2017}\natexlab{}.
\newblock \showarticletitle{PositionRank: An Unsupervised Approach to Keyphrase
  Extraction from Scholarly Documents}. In
  \bibinfo{booktitle}{\emph{Proceedings of the 55th Annual Meeting of the
  Association for Computational Linguistics ({ACL})}}.
\newblock


\bibitem[\protect\citeauthoryear{Grave, Mikolov, Joulin, and Bojanowski}{Grave
  et~al\mbox{.}}{2017}]%
        {fasttext}
\bibfield{author}{\bibinfo{person}{Edouard Grave}, \bibinfo{person}{Tomas
  Mikolov}, \bibinfo{person}{Armand Joulin}, {and} \bibinfo{person}{Piotr
  Bojanowski}.} \bibinfo{year}{2017}\natexlab{}.
\newblock \showarticletitle{Bag of Tricks for Efficient Text Classification}.
  In \bibinfo{booktitle}{\emph{Proceedings of the 15th Conference of the
  European Chapter of the Association for Computational Linguistics,
  {(EACL)}}}.
\newblock


\bibitem[\protect\citeauthoryear{Hamilton, Ying, and Leskovec}{Hamilton
  et~al\mbox{.}}{2017}]%
        {GraphSAGE}
\bibfield{author}{\bibinfo{person}{William~L. Hamilton},
  \bibinfo{person}{Zhitao Ying}, {and} \bibinfo{person}{Jure Leskovec}.}
  \bibinfo{year}{2017}\natexlab{}.
\newblock \showarticletitle{Inductive Representation Learning on Large Graphs}.
  In \bibinfo{booktitle}{\emph{Annual Conference on Neural Information
  Processing Systems (NIPS)}}. \bibinfo{pages}{1024--1034}.
\newblock


\bibitem[\protect\citeauthoryear{Hasan and Ng}{Hasan and Ng}{2014}]%
        {keyprhase_survey}
\bibfield{author}{\bibinfo{person}{Kazi~Saidul Hasan} {and}
  \bibinfo{person}{Vincent Ng}.} \bibinfo{year}{2014}\natexlab{}.
\newblock \showarticletitle{Automatic Keyphrase Extraction: {A} Survey of the
  State of the Art}. In \bibinfo{booktitle}{\emph{Proceedings of the 52nd
  Annual Meeting of the Association for Computational Linguistics ({ACL})}}.
  \bibinfo{pages}{1262--1273}.
\newblock


\bibitem[\protect\citeauthoryear{Hu, Yang, Shi, Ji, and Li}{Hu
  et~al\mbox{.}}{2019}]%
        {GNN_text1}
\bibfield{author}{\bibinfo{person}{Linmei Hu}, \bibinfo{person}{Tianchi Yang},
  \bibinfo{person}{Chuan Shi}, \bibinfo{person}{Houye Ji}, {and}
  \bibinfo{person}{Xiaoli Li}.} \bibinfo{year}{2019}\natexlab{}.
\newblock \showarticletitle{Heterogeneous Graph Attention Networks for
  Semi-supervised Short Text Classification}. In
  \bibinfo{booktitle}{\emph{Proceedings of the 2019 Conference on Empirical
  Methods in Natural Language Processing and the 9th International Joint
  Conference on Natural Language Processing, ({EMNLP-IJCNLP})}}.
  \bibinfo{pages}{4820--4829}.
\newblock


\bibitem[\protect\citeauthoryear{Huang, Ma, Li, Zhang, and Wang}{Huang
  et~al\mbox{.}}{2019}]%
        {GNN_text3}
\bibfield{author}{\bibinfo{person}{Lianzhe Huang}, \bibinfo{person}{Dehong Ma},
  \bibinfo{person}{Sujian Li}, \bibinfo{person}{Xiaodong Zhang}, {and}
  \bibinfo{person}{Houfeng Wang}.} \bibinfo{year}{2019}\natexlab{}.
\newblock \showarticletitle{Text Level Graph Neural Network for Text
  Classification}. In \bibinfo{booktitle}{\emph{Proceedings of the 2019
  Conference on Empirical Methods in Natural Language Processing and the 9th
  International Joint Conference on Natural Language Processing
  ({EMNLP-IJCNLP})}}.
\newblock


\bibitem[\protect\citeauthoryear{Jeh and Widom}{Jeh and Widom}{2002}]%
        {jeh2002simrank}
\bibfield{author}{\bibinfo{person}{Glen Jeh} {and} \bibinfo{person}{Jennifer
  Widom}.} \bibinfo{year}{2002}\natexlab{}.
\newblock \showarticletitle{SimRank: a measure of structural-context
  similarity}. In \bibinfo{booktitle}{\emph{Proceedings of the eighth ACM
  SIGKDD international conference on Knowledge discovery and data mining}}.
  \bibinfo{pages}{538--543}.
\newblock


\bibitem[\protect\citeauthoryear{Kim}{Kim}{2014}]%
        {kim2014convolutional}
\bibfield{author}{\bibinfo{person}{Yoon Kim}.} \bibinfo{year}{2014}\natexlab{}.
\newblock \showarticletitle{Convolutional neural networks for sentence
  classification}.
\newblock \bibinfo{journal}{\emph{arXiv preprint arXiv:1408.5882}}
  (\bibinfo{year}{2014}).
\newblock


\bibitem[\protect\citeauthoryear{Kipf and Welling}{Kipf and Welling}{2017}]%
        {GCN}
\bibfield{author}{\bibinfo{person}{Thomas~N. Kipf} {and} \bibinfo{person}{Max
  Welling}.} \bibinfo{year}{2017}\natexlab{}.
\newblock \showarticletitle{Semi-Supervised Classification with Graph
  Convolutional Networks}. In \bibinfo{booktitle}{\emph{Proceedings of the 5th
  International Conference on Learning Representations ({ICLR})}}.
\newblock


\bibitem[\protect\citeauthoryear{Li, Tang, and Chen}{Li et~al\mbox{.}}{2016}]%
        {Tag_rec}
\bibfield{author}{\bibinfo{person}{Jianguo Li}, \bibinfo{person}{Yong Tang},
  {and} \bibinfo{person}{Jiemin Chen}.} \bibinfo{year}{2016}\natexlab{}.
\newblock \showarticletitle{Leveraging tagging and rating for recommendation:
  {RMF} meets weighted diffusion on tripartite graphs}.
\newblock \bibinfo{journal}{\emph{CoRR}}  \bibinfo{volume}{abs/1611.00812}
  (\bibinfo{year}{2016}).
\newblock


\bibitem[\protect\citeauthoryear{Liu, Chang, Wu, and Yang}{Liu
  et~al\mbox{.}}{2017}]%
        {MTL_text3}
\bibfield{author}{\bibinfo{person}{Jingzhou Liu}, \bibinfo{person}{Wei{-}Cheng
  Chang}, \bibinfo{person}{Yuexin Wu}, {and} \bibinfo{person}{Yiming Yang}.}
  \bibinfo{year}{2017}\natexlab{}.
\newblock \showarticletitle{Deep Learning for Extreme Multi-label Text
  Classification}. In \bibinfo{booktitle}{\emph{Proceedings of the 40th
  International {ACM} {SIGIR} Conference on Research and Development in
  Information Retrieval (SIGIR)}}. \bibinfo{pages}{115--124}.
\newblock


\bibitem[\protect\citeauthoryear{Liu, Qiu, and Huang}{Liu
  et~al\mbox{.}}{2016}]%
        {liu2016recurrent}
\bibfield{author}{\bibinfo{person}{Pengfei Liu}, \bibinfo{person}{Xipeng Qiu},
  {and} \bibinfo{person}{Xuanjing Huang}.} \bibinfo{year}{2016}\natexlab{}.
\newblock \showarticletitle{Recurrent neural network for text classification
  with multi-task learning}.
\newblock \bibinfo{journal}{\emph{arXiv preprint arXiv:1605.05101}}
  (\bibinfo{year}{2016}).
\newblock


\bibitem[\protect\citeauthoryear{Ma, Sun, Yuan, and Cong}{Ma
  et~al\mbox{.}}{2014}]%
        {post_tagging_1}
\bibfield{author}{\bibinfo{person}{Zongyang Ma}, \bibinfo{person}{Aixin Sun},
  \bibinfo{person}{Quan Yuan}, {and} \bibinfo{person}{Gao Cong}.}
  \bibinfo{year}{2014}\natexlab{}.
\newblock \showarticletitle{Tagging Your Tweets: {A} Probabilistic Modeling of
  Hashtag Annotation in Twitter}. In \bibinfo{booktitle}{\emph{Proceedings of
  the 23rd {ACM} International Conference on Conference on Information and
  Knowledge Management ({CIKM})}}. \bibinfo{pages}{999--1008}.
\newblock


\bibitem[\protect\citeauthoryear{Mihalcea and Tarau}{Mihalcea and
  Tarau}{2004}]%
        {TextRank}
\bibfield{author}{\bibinfo{person}{Rada Mihalcea} {and} \bibinfo{person}{Paul
  Tarau}.} \bibinfo{year}{2004}\natexlab{}.
\newblock \showarticletitle{TextRank: Bringing Order into Text}. In
  \bibinfo{booktitle}{\emph{Proceedings of the 2004 Conference on Empirical
  Methods in Natural Language Processing , {EMNLP} 2004, {A} meeting of SIGDAT,
  a Special Interest Group of the ACL, held in conjunction with {ACL} 2004,
  25-26 July 2004, Barcelona, Spain}}. \bibinfo{pages}{404--411}.
\newblock


\bibitem[\protect\citeauthoryear{Papagiannopoulou and
  Tsoumakas}{Papagiannopoulou and Tsoumakas}{2020}]%
        {keyphrase_review}
\bibfield{author}{\bibinfo{person}{Eirini Papagiannopoulou} {and}
  \bibinfo{person}{Grigorios Tsoumakas}.} \bibinfo{year}{2020}\natexlab{}.
\newblock \showarticletitle{A review of keyphrase extraction}.
\newblock \bibinfo{journal}{\emph{Wiley Interdiscip. Rev. Data Min. Knowl.
  Discov.}} \bibinfo{volume}{10}, \bibinfo{number}{2} (\bibinfo{year}{2020}).
\newblock


\bibitem[\protect\citeauthoryear{Sanghoon~Lee and Moon}{Sanghoon~Lee and
  Moon}{2016}]%
        {Tag-IR-survey}
\bibfield{author}{\bibinfo{person}{Janani Balaji Saeid Belkasim
  Rajshekhar~Sunderraman Sanghoon~Lee, Mohamed~Masoud} {and}
  \bibinfo{person}{Seung-Jin Moon}.} \bibinfo{year}{2016}\natexlab{}.
\newblock \showarticletitle{A Survey of Tag-based Information Retrieval}.
\newblock \bibinfo{journal}{\emph{International Journal of Multimedia
  Information Retrieval}}  \bibinfo{volume}{6} (\bibinfo{year}{2016}).
\newblock


\bibitem[\protect\citeauthoryear{Shi, Poghosyan, Ifrim, and Hurley}{Shi
  et~al\mbox{.}}{2018}]%
        {news_tagging_2}
\bibfield{author}{\bibinfo{person}{Bichen Shi}, \bibinfo{person}{Gevorg
  Poghosyan}, \bibinfo{person}{Georgiana Ifrim}, {and} \bibinfo{person}{Neil
  Hurley}.} \bibinfo{year}{2018}\natexlab{}.
\newblock \showarticletitle{Hashtagger+: Efficient High-Coverage Social Tagging
  of Streaming News}.
\newblock \bibinfo{journal}{\emph{{IEEE} Trans. Knowl. Data Eng.}}
  \bibinfo{volume}{30}, \bibinfo{number}{1} (\bibinfo{year}{2018}),
  \bibinfo{pages}{43--58}.
\newblock


\bibitem[\protect\citeauthoryear{Sun, Zhu, Xiao, Xiao, and Wei}{Sun
  et~al\mbox{.}}{2019}]%
        {question_tag_1}
\bibfield{author}{\bibinfo{person}{Bo Sun}, \bibinfo{person}{Yunzong Zhu},
  \bibinfo{person}{Yongkang Xiao}, \bibinfo{person}{Rong Xiao}, {and}
  \bibinfo{person}{Yungang Wei}.} \bibinfo{year}{2019}\natexlab{}.
\newblock \showarticletitle{Automatic Question Tagging with Deep Neural
  Networks}.
\newblock \bibinfo{journal}{\emph{{IEEE Transactions on Learning
  Technologies}}} \bibinfo{volume}{12}, \bibinfo{number}{1}
  (\bibinfo{year}{2019}), \bibinfo{pages}{29--43}.
\newblock


\bibitem[\protect\citeauthoryear{Tang, Yao, Zhang, Xu, Gu, Tong, Yan, and
  Lu}{Tang et~al\mbox{.}}{2019}]%
        {news_tagging_1}
\bibfield{author}{\bibinfo{person}{Shijie Tang}, \bibinfo{person}{Yuan Yao},
  \bibinfo{person}{Suwei Zhang}, \bibinfo{person}{Feng Xu},
  \bibinfo{person}{Tianxiao Gu}, \bibinfo{person}{Hanghang Tong},
  \bibinfo{person}{Xiaohui Yan}, {and} \bibinfo{person}{Jian Lu}.}
  \bibinfo{year}{2019}\natexlab{}.
\newblock \showarticletitle{An Integral Tag Recommendation Model for Textual
  Content}. In \bibinfo{booktitle}{\emph{The Thirty-Third {AAAI} Conference on
  Artificial Intelligence ({AAAI})}}. \bibinfo{pages}{5109--5116}.
\newblock


\bibitem[\protect\citeauthoryear{Veli{\v{c}}kovi{\'c}, Cucurull, Casanova,
  Romero, Lio, and Bengio}{Veli{\v{c}}kovi{\'c} et~al\mbox{.}}{2017}]%
        {velivckovic2017graph}
\bibfield{author}{\bibinfo{person}{Petar Veli{\v{c}}kovi{\'c}},
  \bibinfo{person}{Guillem Cucurull}, \bibinfo{person}{Arantxa Casanova},
  \bibinfo{person}{Adriana Romero}, \bibinfo{person}{Pietro Lio}, {and}
  \bibinfo{person}{Yoshua Bengio}.} \bibinfo{year}{2017}\natexlab{}.
\newblock \showarticletitle{Graph attention networks}.
\newblock \bibinfo{journal}{\emph{arXiv preprint arXiv:1710.10903}}
  (\bibinfo{year}{2017}).
\newblock


\bibitem[\protect\citeauthoryear{Wang, Li, Wang, Zhang, Shen, Zhang, Henao, and
  Carin}{Wang et~al\mbox{.}}{2018}]%
        {MTL_text2}
\bibfield{author}{\bibinfo{person}{Guoyin Wang}, \bibinfo{person}{Chunyuan Li},
  \bibinfo{person}{Wenlin Wang}, \bibinfo{person}{Yizhe Zhang},
  \bibinfo{person}{Dinghan Shen}, \bibinfo{person}{Xinyuan Zhang},
  \bibinfo{person}{Ricardo Henao}, {and} \bibinfo{person}{Lawrence Carin}.}
  \bibinfo{year}{2018}\natexlab{}.
\newblock \showarticletitle{Joint Embedding of Words and Labels for Text
  Classification}. In \bibinfo{booktitle}{\emph{Proceedings of the 56th Annual
  Meeting of the Association for Computational Linguistics ({ACL})}}.
  \bibinfo{pages}{2321--2331}.
\newblock


\bibitem[\protect\citeauthoryear{Wei-Cheng~Chang}{Wei-Cheng~Chang}{2019}]%
        {MTL_text4}
\bibfield{author}{\bibinfo{person}{Kai Zhong Yiming Yang Inderjit~Dhillon
  Wei-Cheng~Chang, Hsiang-Fu~Yu}.} \bibinfo{year}{2019}\natexlab{}.
\newblock \showarticletitle{X-BERT: eXtreme Multi-label Text Classification
  with using Bidirectional Encoder Representations from Transformers}. In
  \bibinfo{booktitle}{\emph{arXiv preprint arXiv:1905.02331}}.
\newblock


\bibitem[\protect\citeauthoryear{Weston, Chopra, and Adams}{Weston
  et~al\mbox{.}}{2014}]%
        {post_tagging_2}
\bibfield{author}{\bibinfo{person}{Jason Weston}, \bibinfo{person}{Sumit
  Chopra}, {and} \bibinfo{person}{Keith Adams}.}
  \bibinfo{year}{2014}\natexlab{}.
\newblock \showarticletitle{{\#}TagSpace: Semantic Embeddings from Hashtags}.
  In \bibinfo{booktitle}{\emph{Proceedings of the 2014 Conference on Empirical
  Methods in Natural Language Processing ({EMNLP})}}.
  \bibinfo{pages}{1822--1827}.
\newblock


\bibitem[\protect\citeauthoryear{Wu, Wu, Li, and Zhou}{Wu
  et~al\mbox{.}}{2016}]%
        {question_tag_3}
\bibfield{author}{\bibinfo{person}{Yu Wu}, \bibinfo{person}{Wei Wu},
  \bibinfo{person}{Zhoujun Li}, {and} \bibinfo{person}{Ming Zhou}.}
  \bibinfo{year}{2016}\natexlab{}.
\newblock \showarticletitle{Improving Recommendation of Tail Tags for Questions
  in Community Question Answering}. In \bibinfo{booktitle}{\emph{Proceedings of
  the Thirtieth {AAAI} Conference on Artificial Intelligence (AAAI)}}.
  \bibinfo{pages}{3066--3072}.
\newblock


\bibitem[\protect\citeauthoryear{Wu, Pan, Chen, Long, Zhang, and Yu}{Wu
  et~al\mbox{.}}{2019}]%
        {GNN_survey}
\bibfield{author}{\bibinfo{person}{Zonghan Wu}, \bibinfo{person}{Shirui Pan},
  \bibinfo{person}{Fengwen Chen}, \bibinfo{person}{Guodong Long},
  \bibinfo{person}{Chengqi Zhang}, {and} \bibinfo{person}{Philip~S. Yu}.}
  \bibinfo{year}{2019}\natexlab{}.
\newblock \showarticletitle{A Comprehensive Survey on Graph Neural Networks}.
\newblock \bibinfo{journal}{\emph{CoRR}}  \bibinfo{volume}{abs/1901.00596}
  (\bibinfo{year}{2019}).
\newblock


\bibitem[\protect\citeauthoryear{Yao, Mao, and Luo}{Yao et~al\mbox{.}}{2019}]%
        {GNN_text2}
\bibfield{author}{\bibinfo{person}{Liang Yao}, \bibinfo{person}{Chengsheng
  Mao}, {and} \bibinfo{person}{Yuan Luo}.} \bibinfo{year}{2019}\natexlab{}.
\newblock \showarticletitle{Graph Convolutional Networks for Text
  Classification}. In \bibinfo{booktitle}{\emph{The Thirty-Third {AAAI}
  Conference on Artificial Intelligence ({AAAI})}}.
  \bibinfo{pages}{7370--7377}.
\newblock


\bibitem[\protect\citeauthoryear{Ying, He, Chen, Eksombatchai, Hamilton, and
  Leskovec}{Ying et~al\mbox{.}}{2018}]%
        {Pinsage}
\bibfield{author}{\bibinfo{person}{Rex Ying}, \bibinfo{person}{Ruining He},
  \bibinfo{person}{Kaifeng Chen}, \bibinfo{person}{Pong Eksombatchai},
  \bibinfo{person}{William~L. Hamilton}, {and} \bibinfo{person}{Jure
  Leskovec}.} \bibinfo{year}{2018}\natexlab{}.
\newblock \showarticletitle{Graph Convolutional Neural Networks for Web-Scale
  Recommender Systems}. In \bibinfo{booktitle}{\emph{Proceedings of the 24th
  {ACM} {SIGKDD} International Conference on Knowledge Discovery {\&} Data
  Mining, ({KDD})}}. \bibinfo{pages}{974--983}.
\newblock


\bibitem[\protect\citeauthoryear{Zhang and Zhou}{Zhang and Zhou}{2014}]%
        {MTL_survey}
\bibfield{author}{\bibinfo{person}{Min{-}Ling Zhang} {and}
  \bibinfo{person}{Zhi{-}Hua Zhou}.} \bibinfo{year}{2014}\natexlab{}.
\newblock \showarticletitle{A Review on Multi-Label Learning Algorithms}.
\newblock \bibinfo{journal}{\emph{{IEEE} Trans. Knowl. Data Eng.}}
  \bibinfo{volume}{26}, \bibinfo{number}{8} (\bibinfo{year}{2014}),
  \bibinfo{pages}{1819--1837}.
\newblock


\bibitem[\protect\citeauthoryear{Zheng and Casari}{Zheng and Casari}{2018}]%
        {zheng2018feature}
\bibfield{author}{\bibinfo{person}{Alice Zheng} {and} \bibinfo{person}{Amanda
  Casari}.} \bibinfo{year}{2018}\natexlab{}.
\newblock \bibinfo{booktitle}{\emph{Feature engineering for machine learning:
  principles and techniques for data scientists}}.
\newblock \bibinfo{publisher}{" O'Reilly Media, Inc."}.
\newblock


\bibitem[\protect\citeauthoryear{Zhou, Gou, Hu, Zhang, Xu, Jiang, Li, and
  Xiong}{Zhou et~al\mbox{.}}{2019}]%
        {POI_tagging}
\bibfield{author}{\bibinfo{person}{Jingbo Zhou}, \bibinfo{person}{Shan Gou},
  \bibinfo{person}{Renjun Hu}, \bibinfo{person}{Dongxiang Zhang},
  \bibinfo{person}{Jin Xu}, \bibinfo{person}{Airong Jiang},
  \bibinfo{person}{Ying Li}, {and} \bibinfo{person}{Hui Xiong}.}
  \bibinfo{year}{2019}\natexlab{}.
\newblock \showarticletitle{A Collaborative Learning Framework to Tag
  Refinement for Points of Interest}. In \bibinfo{booktitle}{\emph{Proceedings
  of the 25th {ACM} {SIGKDD} International Conference on Knowledge Discovery
  {\&} Data Mining ({KDD})}}. \bibinfo{pages}{1752--1761}.
\newblock


\bibitem[\protect\citeauthoryear{Zhu, Li, Ouyang, Yu, and Wang}{Zhu
  et~al\mbox{.}}{2017}]%
        {image_tagging_3}
\bibfield{author}{\bibinfo{person}{Feng Zhu}, \bibinfo{person}{Hongsheng Li},
  \bibinfo{person}{Wanli Ouyang}, \bibinfo{person}{Nenghai Yu}, {and}
  \bibinfo{person}{Xiaogang Wang}.} \bibinfo{year}{2017}\natexlab{}.
\newblock \showarticletitle{Learning Spatial Regularization with Image-Level
  Supervisions for Multi-label Image Classification}. In
  \bibinfo{booktitle}{\emph{2017 {IEEE} Conference on Computer Vision and
  Pattern Recognition, {CVPR} 2017, Honolulu, HI, USA, July 21-26, 2017}}.
  \bibinfo{publisher}{{IEEE} Computer Society}, \bibinfo{pages}{2027--2036}.
\newblock
\urldef\tempurl%
\url{https://doi.org/10.1109/CVPR.2017.219}
\showDOI{\tempurl}


\end{thebibliography}
\end{document}